\documentclass{llncs}
\usepackage{makeidx}  
\usepackage{graphicx}
\usepackage{listings}

\usepackage{cite}
\begin{document}
\frontmatter          
\pagestyle{headings}  
\addtocmark{Feature-diverse Test Data} 
%

%
%
\mainmatter              
%
\title{Searching for test data with feature diversity}
\titlerunning{Feature-diverse Test Data}  
%
\author{Robert Feldt and Simon Poulding}
\authorrunning{Authors suppressed for blind review} 
%
\tocauthor{Authors/institutions suppressed for blind review}
\institute{Chalmers University of Technology and Blekinge Institute of Technology\\
\email{robert.feldt@chalmers.se, robert.feldt@bth.se},\\ WWW home page:
\texttt{http://www.robertfeldt.net}
}

\maketitle              

\begin{abstract}
There is an implicit assumption in software testing that more diverse and varied test data is needed for effective testing and to achieve different types and levels of coverage.
Generic approaches based on information theory to measure and thus, implicitly, to create diverse data have also been proposed.
However, if the tester is able to identify features of the test data that are important for the particular domain or context in which the testing is being performed, the use of generic diversity measures such as this may not be sufficient nor efficient for creating test inputs that show diversity in terms of these features.
Here we investigate different approaches to find data that are diverse according to a specific set of features, such as length, depth of recursion etc.
Even though these features will be less general than measures based on information theory, their use may provide a tester with more direct control over the type of diversity that is present in the test data.
Our experiments are carried out in the context of a general test data generation framework that can generate both numerical and highly structured data.
We compare random sampling for feature-diversity to different approaches based on search and find a hill climbing search to be efficient. The experiments highlight many trade-offs that needs to be taken into account when searching for diversity.
We argue that recurrent test data generation motivates building statistical models that can then help to more quickly achieve feature diversity.

\end{abstract}

\section{Introduction}
Most testing practitioners know that a key to high-quality testing is to use diverse test data. However, it is only recently that there has been research to formalise different notions of diversity and propose concrete metrics to help realise it~\cite{feldt2008searching, alshahwan2012augmenting, feldt2016testsetdiameter, shi2016}.
The diversity that is sought is often of a general shape and form, i.e.\ rather than target some specific attribute or feature of the test data we seek diversity in general.
Even though this is appropriate when little is known about the test data that is needed, it makes it harder for testers to judge if diversity has really been achieved and of which type.
Moreover, if the tester has some prior information or preference as to which type of test data to explore it is not clear, in the general diversity context, how to incorporate this during testing.

Here we target a specific form of test diversity (TD) that we call the Feature-Specific TD problem: how to sample as diverse and complete set of test inputs as possible in a specific area of the feature space. As a concrete example, for software-under-test that takes strings as inputs, a tester might prefer test inputs that are in a particular size range (feature 1)  and for which the count of numeric characters is within a given range (feature 2).
This problem is in contrast to the General TD problem where we seek diversity in general without requiring diversity within in a particular set of features.

There is existing work on how to search for test data with one specific set of feature values~\cite{feldt2013finding}, 
as well as techniques that address the General TD problem, but there is a lack of work on the Feature-Specific TD problem.
In this paper we propose a variety of methods to generate test data with specific types of feature diversity, and then explore their strengths and weaknesses in order to understand the trade-offs between them.


Our contributions are:

\begin{itemize}
    \item Identification of multiple basic methods to search for feature-specific test diversity,
    \item Evaluation of the basic approaches on a two-dimensional feature space for a test data generation problem,
    \item Proposal of hybrid search methods based on the results of the evaluation.
\end{itemize}

%
%

%
%

In Section~\ref{sec:Background} we provide further background and summarise related work. In Section~\ref{sec:Design} we propose the search-based methods to seek feature-specific diversity, and describe and discuss their evaluation in Section~\ref{sec:Evaluation}. 
We then summarise our conclusions in Section~\ref{sec:Conclusions}.

\section{Background and Related Work}
\label{sec:Background}

Test data generation techniques often apply a strategy that has an implicit objective of ensuring some form of diversity in the set of test inputs that are created. Testing techniques that partition the input domain -- for example, based on the structural coverage of the software-under-test -- select a representative test input from each partition and so implicitly achieve diversity in the context of the criterion used for partitioning.  Even uniform random testing implicitly achieves some form of diversity simply because every input in the input domain has the same, non-zero probability of being selected for the test set.

In contrast, there exist test data generation techniques that have an \emph{explicit} objective of diversity within the input domain.  One class of such techniques use a distance metric between two test inputs, such as the Euclidean distance between numeric inputs, and interpret this metric as a measure of diversity to guide the selection of test inputs.  

Antirandom Testing chooses a new test input such that is maximises the total distance between the new datum and all the existing inputs already in the test set \cite{malaiya1995antirandom}.  Adaptive Random Testing first creates a pool of candidate inputs by random selection, and then adds to the test set the input in the candidate pool for which the minimum distance from all existing members of the test set is the largest \cite{chen2004adaptive}.  Both these techniques therefore create a set of test inputs element-by-element by selecting the next test element to be as dissimilar as possible from existing elements.

Bueno et al.\ consider instead the set of test inputs as a whole, and define a diversity metric on the set as sum of the distances from each input to it nearest neighbour \cite{bueno2007improving}. Metaheuristic search is then applied to the set of test inputs with the objective of maximising the diversity metric. Hemmati et al.\ apply both the element-wise approach of Adaptive Random Testing and a whole-set approach similar to that of Bueno et al.\ to the selection of diverse cases derived using model-based testing \cite{hemmati2011empirical}.

Diversity metrics based on Euclidean distance are limited in terms of the types of inputs to which they can be applied.  Feldt et al.\ demonstrate that normalised compression distance, a distance metric based on information theory, is not limited in the data types to which it may be applied, and enables the selection test inputs in a manner similar to how a human would based on 'cognitive' diversity \cite{feldt2008searching}.  Normalised compression distance is a pair-wise metric, but a recent advance in information theory extends this notion to a set as a whole.  Feldt et al.\ use this set-wise metric to introduce test set diameter, a diversity metric that is applied to the entire test set, and demonstrate how this metric can be used to create diverse test sets \cite{feldt2016testsetdiameter}.

Panichella et al. demonstrate an alternative mechanism for promoting diversity in the context of selecting test cases for regression testing.  Instead of a search objective based on diversity, the authors propose a multi-objective genetic algorithm in which the genetic operators -- in this case the initialisation of the population and the generation of new individuals -- are designed to `inject' diversity at the genome level \cite{panichella2015improving}.

In the above approaches, the notion of diversity is generic in the sense that it is agnostic as to the `meaning' of the test inputs.  Metrics such as Euclidean distance or normalised compression distance simply treat the inputs as numeric vectors or strings of symbols, respectively, rather than aircraft velocities, time-series of temperature measurements, or customer addresses etc. The advantage of generic diversity metric and generic algorithm operators is that they can be applied easily to any domain, but the risk is that they may overlook domain-specific notions of diversity that might be important in deriving effective test inputs.

In this paper, we investigate instead how to measure and apply diversity that takes into account the domain-specific meaning of the test inputs.  We take inspiration from a recent class of evolutionary algorithms known as illumination algorithms or quality diversity algorithms \cite{pugh2016quality}.  These algorithms differ from traditional evolutionary algorithms in that they forego the use of objective fitness as the primary pressure that drives the selection of new individuals, and instead select new individuals based on the domain-specific `novelty' of the phenotype. The premise is that the search for novelty maintains diversity and avoids premature convergence to local optimum.  Or considered another way, the pressure for ever-increasing objective fitness can prevent the algorithm from finding the sequence of `stepping stones' that leads to the glabal optimum. These algorithms have been shown find near-globally optimum solutions as a by-product of the search for novelty.

For example, Lehman and Stanley's novelty search algorithm evaluates new individuals in terms of a novelty metric, and this metric is unrelated to the objective metric  \cite{lehman2008exploiting}.  To calculate the novelty metric, domain-specific features of the phenotype are measured to obtain a feature vector, and then measures the distance of the individual from its nearest neighbouring individuals in this feature space: the larger this distance, the more novel the individual is considered to be.

The Multi-dimensional Archive of Phenotypic Elites (MAP-Elites) algorithm of Mouret and Clune uses the feature-space to maintain diversity in a different manner: at each point in the feature space (which is discretised for this purpose), an archive is maintained of the best individual having the features, where best is measured in terms of objective fitness \cite{mouret2015illuminating}.  The set of these elite individuals -- one at each point in the feature space -- is the population on which the evolutionary algorithm acts.

We note that Marculescu et al.\ apply both novelty search and MAP-Elites to generate candidate test inputs as part of an interactive search-based software testing system, and found that, compared to a traditional objective-based evolutionary algorithm, the illumination algorithms found more diverse test cases \cite{marculescu2016using}.  

It is this general strategy of illumination algorithms -- that of searching for diversity in a domain-specific feature space -- that informs the work in this paper.  In addition, the specific strategies employed by Novelty Search and MAP-Elites are the basis for some the approaches we investigate.






\section{Focused Search for Feature Diversity}
\label{sec:Design}

The research described in this paper is motivated by the premise that test data chosen for feature-specific test diversity will be more effective than test data chosen according to more generic measures of diversity (such as those discussed in section~\ref{sec:Background} above).  The objective of the research is then to explore a number of search-based methods for choosing test inputs with high feature-specific diversity.

In this section, we describe:
\begin{itemize}
    \item a concrete testing scenario (described first since a feature space is scenario-specific);
    \item a feature space for the testing scenario;
    \item a base mechanism for generating test inputs for this scenario;
    \item a set of search-based methods that can be applied to base mechanism to promote feature-specific diversity (the empirical work in section~\ref{sec:Evaluation} will compare the effectiveness and efficiency of these methods).
\end{itemize}

\subsection{Testing Scenario}
The input domain consists of strings that are arithmetic expressions formed from the operators \verb|+|, \verb|-|, \verb|*|, and \verb|/|; integers; and parentheses. An example of a valid input is the string: \verb|"42+(-7*910)"|.  We choose this domain since it is realistically complex: inputs are not simply numeric, but instead a string of characters that must satisfy constraints on its structure, and there is no bound on the length of the expression string.

We do not explicitly define the software-under-test in this scenario since the search-based methods we apply act on the inputs themselves rather than on the coverage or other information from executing the software. But we have in mind software that parses the arithmetic expression and calculates the result.

\subsection{Feature Space}
By `feature space', we mean the specification of one or more named dimensions on which test inputs can vary on a defined scale. Typically this scale is numerical and each feature has associated with it a specific function that maps input onto the scale, but the scale can also be ordinal or categorical

For the purposes of the empirical work, we consider a two-dimensional feature space formed by:
\begin{description}
    \item[Feature 1: Length] -- the number of characters in the string
    \item[Feature 2: NumDigits] -- the number of characters that are digits (`0' to `9' in the ASCII range)
\end{description}

We envision that the tester wishes to generate a large number of test inputs that differ in both total length as well as in the number of digits. 

Feature spaces can be very large, and may be infinite.  This is indeed the case in this scenario: there is no bound on the length of either the expression string, nor the number the number of digits in the string.  Therefore a tester needs to define a preferred area of the feature space where testing should be focused.  For example, she may specify a range of values for each feature that together define a hypercube within the feature space. 


\subsection{Base Generation Mechanism}
In order to generate valid test inputs, we use Feldt and Poulding's G\"{o}delTest framework for generating structured data \cite{feldt2013finding}.  In this framework, a programmatic generator is used to define the structure of valid inputs -- here, the structure of valid arithmetic expressions
\footnote{The generator we use for arithmetic expressions is the same as that included as an example in the README file for the DataGenerators package at: \url{https://github.com/simonpoulding/DataGenerators.jl}.  The DataGenerators package is Feldt and Poulding's implementation of G\"{o}delTest in the language Julia.}
-- and a choice model is used to control which of all the possible valid arithmetic expressions is emitted by the generator.

For this work, we use stochastic choice models that, in effect, define a probability distribution over the space of all valid arithmetic expressions.  With such choice models, G\"{o}delTest becomes a mechanism for generating random arithmetic expressions according to the distribution defined by the choice model. 

Choice models in G\"{o}delTest have parameters that can be used to change the probability distribution, and the search-based methods for diversity described below operate by manipulating these parameters.

The empirical work considers two stochastic choice models:
\begin{description}
    \item[Default] The default `sampler' choice model provided by G\"{o}delTest.  When used with the arithmetic expression generator, this choice model has 8 parameters all in the range $[0.0, 1.0]$.
    \item[RecDepth5] An extension of the default choice model that enables more refined probability distribution.  Specifically, the probabilistic choice of whether an operand in the expression is a number, or is itself a parenthesised subexpression, becomes conditional on the depth to which the current subexpression is nested.  This choice model has 16 parameters, again all in the range $[0.0, 1.0]$.
\end{description}

\subsection{Search-Based Methods}
Our goal is to cover as large a portion as possible of the preferred area in the feature space. The fundamental approach we take is based on the novelty search algorithm described in section~\ref{sec:Background} above.  The density (or simpler, even the count) of test inputs in a specific cell of the preferred area of the feature space as metric is used to guide the search to areas with lower density so that novel inputs can be found that will improve diversity of the test set as a whole. In addition, we consider several types of random search as baselines and investigate a more expressive stochastic model to govern the sampling of test inputs.

For random sampling, one can either set the G\"{o}delTest choice model parameters (which define a probability distribution over the valid inputs) to random values (i.e.\ to define a distribution at random) once at the start of the generation process, or continuously during the process. We call the former method \texttt{rand-once} and designate the latter \texttt{rand-freqN} with N denoting the frequency with which we resample the parameters. From previous research, it is known that some stochastic choice models can be quite brittle and lead to large numbers of `infeasible' inputs -- inputs that are extremely large or infinite and exceed the finite memory available to represent them -- being generated.  For this reason we also include a \texttt{rand-mfreqN} method denoting up to a maximum of N inputs sampled between resampling events. The maximum means that as soon as an infeasible input is generated, we directly resample random values for the choice model parameters.

For random sampling it is well-known that so called Latin Hypercube Sampling (LHS) can generate a better `spread' of samples over a space~\cite{park1994optimal}. When using LHS one first divides the value range for each dimension being sampled into equal-sized bins and then samples within each bin. This ensures that each dimension is sampled over the full range of its values. 
We select 10 and 30 bins respectively and designate the corresponding methods \texttt{rand-mfreq5-LHS10} and \texttt{rand-mfreq10-LHS30}.

We also include Nested Monte-Carlo Search (NMCS) \cite{cazenave2009nested}, a form of Monte-Carlo Tree Search that has been previously applied successfully to guide the generation of test inputs by G\"{o}delTest~\cite{poulding2014}.  NMCS operates during the generation process itself rather than on the choice model parameters.  Each time a decision needs to be made -- such as whether an operand in the arithmetic expression is a number or a subexpression, or the number of digits in a number operand -- NMCS performs an internal `simulation' by taking each possible choice for that decision in turn, and for each, then completing the generation process as normal (i.e.\ using choice determined by the choice model).  Whichever simulation results in the best outcome, the corresponding choice is made for that decision.

The variant of NMCS used by G\"{o}delTest considers a fixed sample of possible choices, rather than all possible choices, since there may be infinite number of such choices for a decision.  We consider two variants in the evaluation: one that uses a sample of 2 choices at each decision, and the other uses a sample of 4 choices.

NMCS generates many `intermediate' candidate test inputs as outcomes from its internal simulations, and there are several options for utilising these intermediates. We argue that it makes sense to not throw away these intermediates but rather use them to update the density used in fitness calculations. We include both an approach that updates the density directly and thus changes the fitness calculation for all subsequent samples, and a batch approach that fixes density during one exploration by the NMCS algorithm and then uses all intermediate test inputs to update the density in one go before the next generation is started.

Thus, we use four NMCS methods in the empirical evalution: \texttt{nmcs-2-direct}, \texttt{nmcs-4-direct}, \texttt{nmcs-2-batch}, and \texttt{nmcs-4-batch}.

Finally, we include a hill climbing method that is applied to parameters of the choice model.  Since this is not a population-based method it is easier to control in detail how it compares the diversity of the inputs generated by new candidate parameters to the current model parameters. A new candidate is formed by making small changes to the current model parameters using a Gaussian distribution with a small standard deviation.  We adapt the sampling and comparison step used in a traditional hill climber to try to minimize the number of sampled test inputs. After sampling a minimum number of inputs (4) we sample up to a maximum number (20) while discarding the new point if it generates more than 33\% infeasible inputs, or 50\% feasible inputs that are outside the preferred area of the feature space. It uses a Mann-Whitney U test to compare the densities in feature space of the test inputs sampled by the current parameters and the new parameters, and goes to the latter if the $p$-value of the test is below 20\%. The settings (the number of samples, $p$-value threshold, etc.) were chosen in an ad hoc manner, but the method seemed robust to changes in them during initial testing so we did not tune them further. This method is denoted \texttt{hillclimb-4-20}.



\section{Empirical Evaluation}
\label{sec:Evaluation}

We applied all 10 methods defined above to search for diverse test inputs in the two-dimensional feature space defined by the string length and number of digits of the input. Each method was executed 25 times\footnote{Except for one long-running method, as detailed later.} to account for (stochastic) variation in their performance. Below we discuss the results from two different perspectives: the coverage of the preferred area of the feature space, and the efficiency of the methods, i.e. their coverage compared to the search time they needed.


\subsection{Feature space coverage}

Once a tester has defined a particular preferred area of a feature space where she is interested in focusing attention our main concern is to create a set of test inputs that cover this area to the largest extent possible. Although there often exists many constraints between features, a tester may rarely be aware of them or have the time to define them in detail. We will thus assume that the focus area has the shape of a hypercube in the feature space, i.e. its limits are defined with one or more ranges of preferred values per feature. In this context it is natural to consider coverage in terms of how many of the unique combinations of feature values that has been covered during a search.

For example, in the two-dimensional feature space used here we have used the preferred area of lengths between 3~\footnote{We did not start at length 1 since the grammar of this particular generator is specified such that the shortest string possible is the one with two single-digit numbers with a single, binary, numeric operator between them, for a minimum string length of 3.} and 50 and number of digits between 2 and 25 (both ranges inclusive). There is clearly a constraint between the features here, the number of digits has to be smaller than equal to the length of the string, but for other feature spaces and preferred areas the effect of dependencies and constraints might be harder to identify. We thus will use the theoretical maximum size of the \textit{preference hypercube}
We call this \textit{Feature Space Hypercube Coverage} (FSHC), and denoted simply coverage in the following.

By definition this means that very rarely can a search method even in theory reach 100\% FSHC for a specific preference hypercube; FSHC values should be compared only in relation to each other and not on an absolute scale. For a specific feature space, preference hypercube and set of searches to fill it one can normalize the FSHC by the largest FSHC value seen and thus calculate the \textit{Normalized Feature Space Hypercube Coverage} (NFSHC). In the example we have used here the size of the preference hypercube is $(50-3+1)*(25-2+1)$ which is $1152$. The largest number of unique feature vectors cells covered, in a (long-running) search using Hill Climbing, was 651, which means that the largest FSHC observed in our experiments was $56.5\%$.

A summary of the overall performance of the 10 different methods we investigate can be seen in Table~\ref{tab:descriptive_stats}. The number of runs per method was 25 except for \texttt{rand-freq1} where we limited the number of repetitions due to the longer search time.

\begin{table}[p]
\begin{center}
\begin{tabular}{ccccccc}
\hline
Method & ChoiceModel & Runs & Coverage & std & Time & Preferred\\
\hline
$\tt{hillclimb-4-20}$ & $\tt{RecDepth5}$ & $25$ & $\tt{52.7}$ & $\tt{1.3}$ & $\tt{235.9}$ & $\tt{80.5}$\\
$\tt{rand-mfreq5-LHS10}$ & $\tt{RecDepth5}$ & $25$ & $\tt{52.5}$ & $\tt{0.5}$ & $\tt{519.4}$ & $\tt{65.7}$\\
$\tt{rand-mfreq10-LHS30}$ & $\tt{RecDepth5}$ & $25$ & $\tt{52.3}$ & $\tt{0.5}$ & $\tt{348.7}$ & $\tt{66.8}$\\
$\tt{rand-freq1}$ & $\tt{RecDepth5}$ & $25$ & $\tt{52.2}$ & $\tt{0.5}$ & $\tt{980.1}$ & $\tt{61.9}$\\
$\tt{rand-freq1}$ & $\tt{Default}$ & $10$ & $\tt{49.1}$ & $\tt{0.8}$ & $\tt{2237.1}$ & $\tt{51.1}$\\
$\tt{nmcs-4-direct}$ & $\tt{Default}$ & $25$ & $\tt{46.4}$ & $\tt{1.6}$ & $\tt{217.6}$ & $\tt{62.4}$\\
$\tt{nmcs-2-direct}$ & $\tt{Default}$ & $25$ & $\tt{45.4}$ & $\tt{1.2}$ & $\tt{231.3}$ & $\tt{61.9}$\\
$\tt{nmcs-2-batch}$ & $\tt{Default}$ & $25$ & $\tt{45.2}$ & $\tt{1.2}$ & $\tt{234.3}$ & $\tt{61.5}$\\
$\tt{nmcs-4-batch}$ & $\tt{Default}$ & $25$ & $\tt{44.7}$ & $\tt{1.2}$ & $\tt{228.6}$ & $\tt{61.7}$\\
$\tt{rand-once}$ & $\tt{Default}$ & $25$ & $\tt{39.6}$ & $\tt{0.4}$ & $\tt{265.2}$ & $\tt{64.0}$
\end{tabular}

\caption{Descriptive statistics on the performance of the 10 investigated methods on the 2-dimensional feature space of string length and number of digits for the ExprGen generator. The `Runs' columns shows the number of runs per method, `Coverage' shows the mean FSHC while `std' is its standard deviation. Finally, `Time' is the mean search time in seconds and `Preferred' is the ratio of samples that is within the preference hypercube.} 
\label{tab:descriptive_stats} 
\end{center}
\end{table}

From the table we see that Hill Climbing performs well but the methods based on random resampling have competitive performance and reach similar levels of coverage. We also see that the NMCS-based methods are generally fast but have worse coverage, regardless if using direct or batch updating of the density. It is also clear that a major determinant of coverage, in addition to the method used, is the choice model. All the methods using the default sampler choice model are at the bottom of the table based on the mean FSHC level reached. A striking example of the difference the choice model can make is for the \texttt{rand-freq1} method which reaches an average FSHC of $52.2\%$ with the recursive model at depth 5 while it only reaches $49.1\%$ with the default sampler choice model. This is a statistically significant difference with a p-value less than $0.00001$ based on a Mann-Whitney U-test.



\begin{figure}[htp]
\centering
\includegraphics[width=9cm]{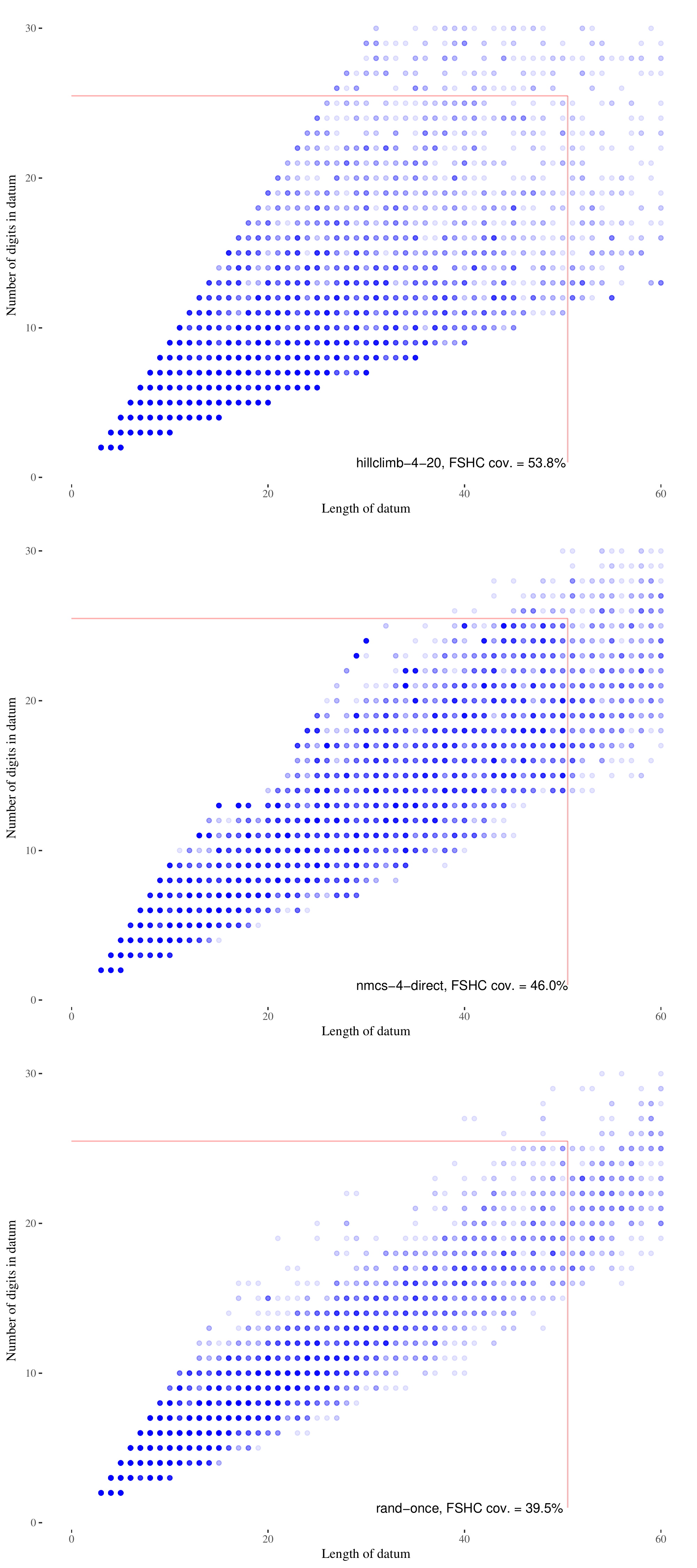}
\caption{Scatterplots showing the actual coverage of the preference hypercube (its upper limit is marked with red lines) after 10,000 data were sampled by three different methods: \texttt{hillclimb-4-20} (top), \texttt{nmcs-4-direct} (middle), and \texttt{rand-once} (bottom).}
\label{fig:featurespace_plots}
\end{figure}

An advantage of using a two-dimensional feature space is that we can visualise in more detail the test data diversity of found by the methods. Figure~\ref{fig:featurespace_plots} shows scatterplots for one run each of the three methods \texttt{hillclimb-4-20} (top), \texttt{nmcs-4-direct} (middle), and \texttt{rand-once} (bottom). These plots draw one point per found test input using a low alpha (transparency) value; thus the darkness of the dot in each cell gives an indication of the density with which the cell was covered. We can see the superior coverage of hill climbing that manages to cover also cells of the hypercube on top where the length of the string (x axis) has medium to low values while the number of digits (y axis) is as high as possible. We can also see that the NMCS search, the middle graph, seems to be constrained in a similar way as the random once method in the bottom graph, i.e. they both have problems to cover the upper parts of the preference hypercube. The NMCS methods are constrained by the base sampler choice model use for the internal simulations.

\subsection{Efficiency - Coverage per time}

\begin{figure}[htp]
\centering
\includegraphics[width=11.7cm]{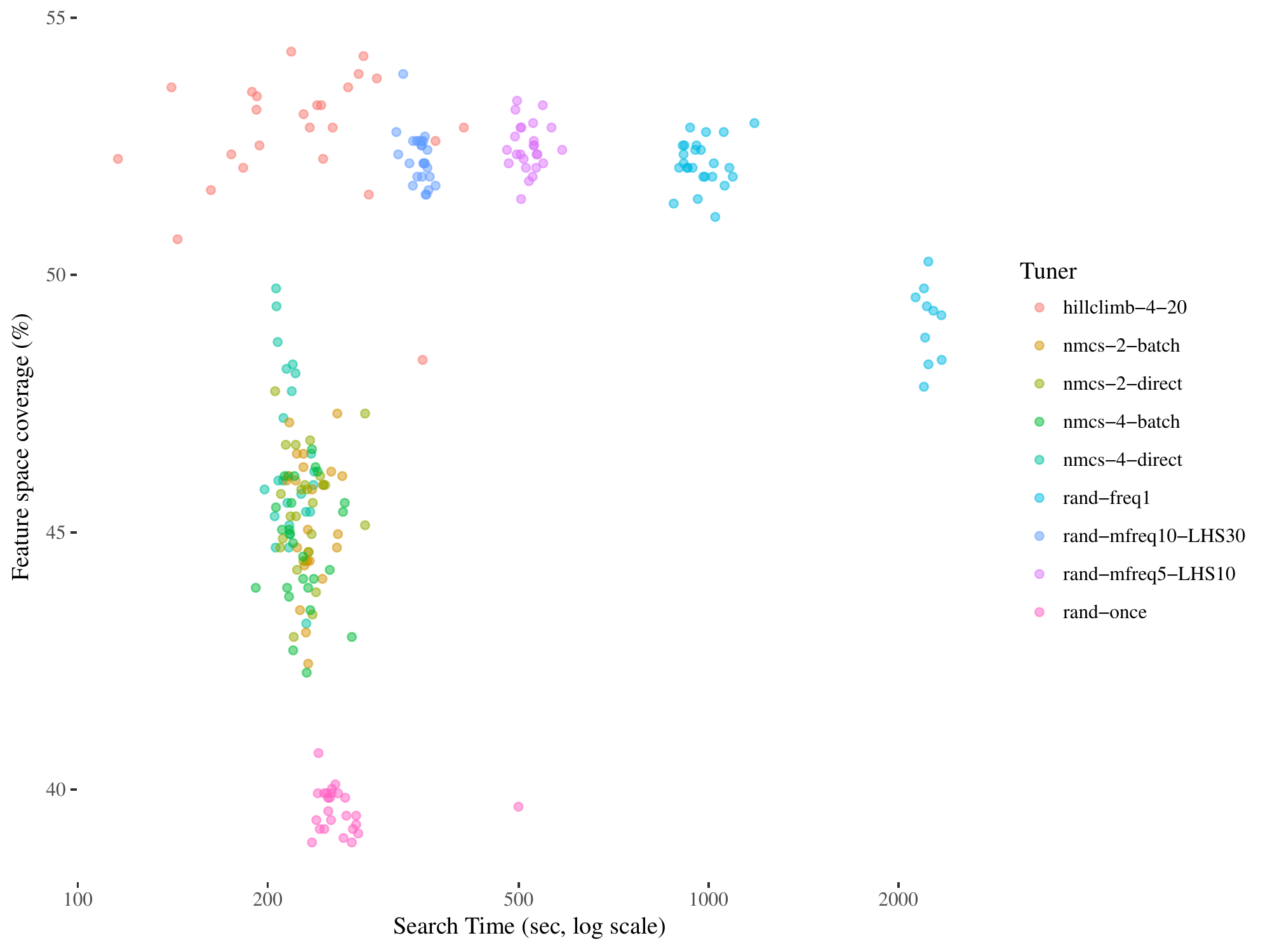}
\caption{Feature space coverage (in \% of theoretical maximal feature space size) versus the search time (in seconds) used by a method to reach that coverage. The scale on the X axis is logarithmic}
\label{fig:coverage_vs_time}
\end{figure}

To study the overall efficiency of the tested methods in more detail we can plot the coverage level reached versus the search time expended. Figure~\ref{fig:coverage_vs_time} shows a scatterplot with the search time in seconds on the (logarithmically scaled) X axis, and the percent of preferred feature space covered (FSHC) on the Y axis. The colour of each point in the graph codes for the method used, so a cloud of points of the same colour represents all the runs of one and the same method. The best position in this graph would be up and on the left meaning a run that both got a high coverage and had a low search time.

Consistent with the results shown previously we can see that the Hill Climbing method has consistently good results. Even if its variance is larger than for the other methods, signified by the `lone' light orange dot towards the middle of the graph, it tends to be among the fastest optimisers while also reaching the highest coverage levels, on average.

This can be contrasted with the simplest possible strategy, \texttt{rand-once}, in light pink down at the bottom of the graph. Even if, on average, it has similar run times to the fastest methods it fails to even reach 40\% coverage.

The NMCS methods all use the default choice model for sampling while traversing the tree of choices. Thus it seems to be hampered by the low coverage of this base model. Even though the NMCS search seems to be able to `push out' from the confines of its base stochastic model, and thus each higher levels of coverage it does not reach as high as the Hill Climber or the methods based on random (re-)sampling of parameters. This makes it clear that NMCS needs a good base model adapted to the task. Alternatively it hints at the possibility to hybridize NMCS by dynamically adapting or randomly sampling the underlying stochastic model.

Figure~\ref{fig:coverage_vs_time} also gives us an opportunity to better understand what causes long search times for a method. If we look at the three middle point clouds on top, for \texttt{rand-freq1} (light green, middle right), \texttt{rand-mfreq5-LHS10} (light purple, middle), and \texttt{rand-mfreq10-LHS30} (light blue, middle left), we see that they reach roughly the same coverage levels. However, the \texttt{mfreq10} version takes less than half the search time of the \texttt{freq1} version to reach that coverage. Since the maxfreq construct will resample a new set of parameter values early if an infeasible datum is generated, time tends to be saved. This is since the infeasible inputs typically arise when the stochastic model is configured to lead to a deep recursion in the number of method calls.

\begin{figure}[htp]
\centering
\includegraphics[width=10.7cm]{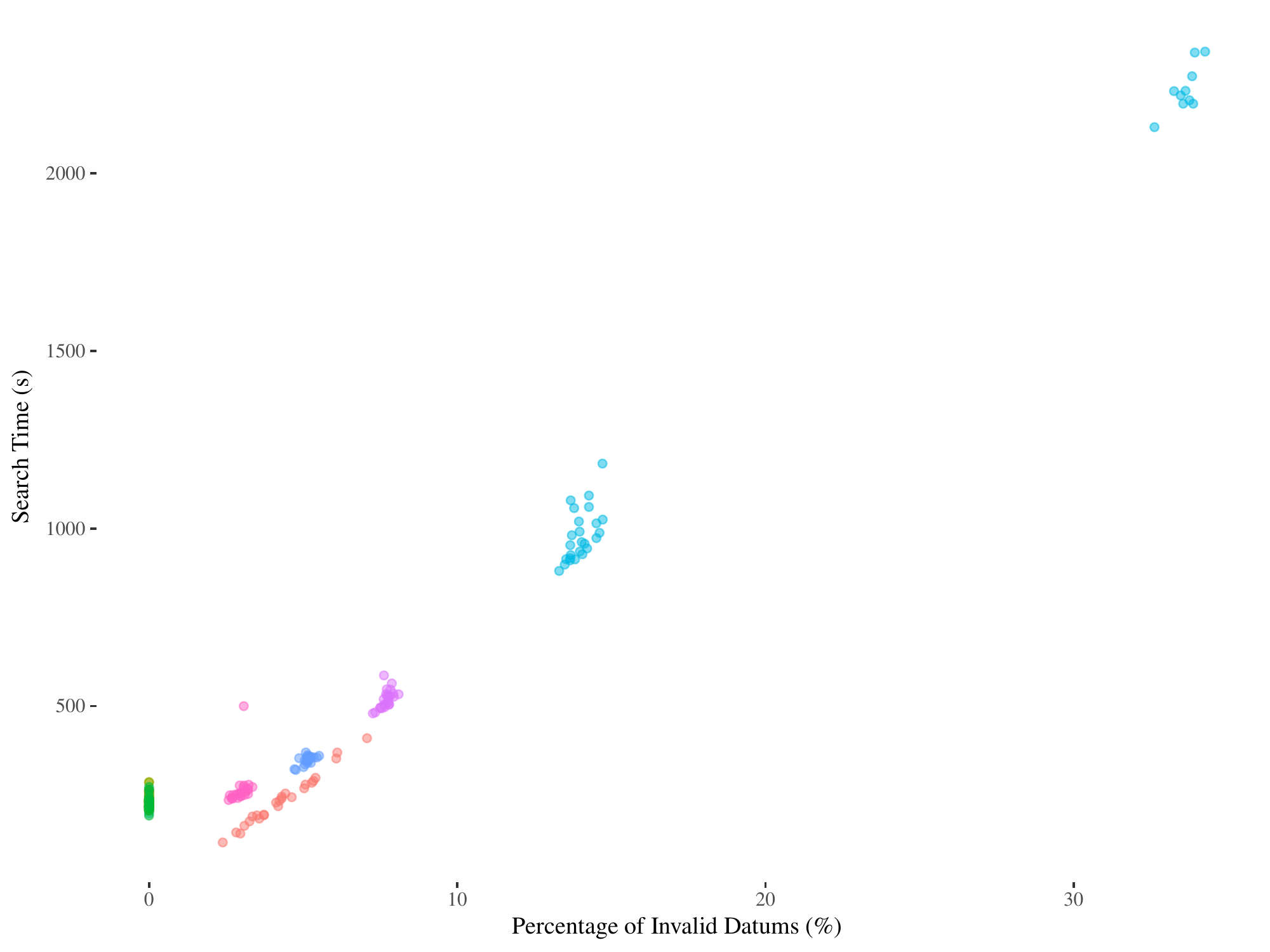}
\caption{Search time (in seconds) used by a run versus the percentage of generated test inputs that are infeasible. The colour of the points in the graph are the same as in Figure~\ref{fig:coverage_vs_time} above so legend excluded here.}
\label{fig:invalid_vs_time}
\end{figure}

We can see this effect more clearly if we plot the search time for each run versus the percentage of infeasible values sampled during the run. Figure~\ref{fig:invalid_vs_time} shows that, except for the NMCS methods on the left, there is an almost linear relation between these factors. The smaller but still additional search time increase seen for the rightmost runs in each cluster is probably from the fact that before a deep recursion during generation is interrupted, and an infeasible value returned, there is a large space of non-preferred but still feasible part of the feature space. Generating such test inputs will also take longer than generating shorter inputs with a few levels of recursion. The only real exception to strong correlation seen are the NMCS methods which have a close to 0\% of sampled inputs being invalid and still having a relatively high search time. The nature of the NMCS search process is that as soon as one non-preferred datum is generated during the tree-wise `pruning' of choices the whole sub-tree of choices will be deselected; and sub-sequent choices are thus less likely to lead to non-referred or infeasible data.

\subsection{Discussion}
\label{sec:Discussion}


Through a set of experiments with 10 different methods to generate diverse test data in specific areas of a defined feature space we have shown that there is not one clearly better method to employ. The results show that a simple hill climbing search was relatively more efficient in covering the preferred parts of the feature space: it covered a larger part of the area in less time. However, random alternatives were not far behind and offer alternative benefits such as less bias. With any search algorithm there is always the risk that one is trading efficiency on one particular set of problems with efficiency in general, over all problems (see for example the `No Free Lunch' theorems by Wolpert and Macready~\cite{wolpert1997}). This can be problematic if the bias leads to the tester missing erroneous behavior of the software under test. However, if a tester really has a reason to want to target a smaller area of the input space a more directed search, such as using a hill climbing search, can be called for.



An important finding in our experiments is about the test data generation tool itself. Even though the Nested Monte Carlo Search (NMCS) has been previously shown by Poulding and Feldt~\cite{poulding2014} to better target test data with very specific features when we here tried their approach to cover a feature space it is clear that NMCS can be hampered by its underlying stochastic model.
All methods we evaluated consistently performed better when using a larger than default stochastic model that gives more detailed control of the generation process. 
Such models allow for more fine-grained control that can be exploited by the searchers but also used for more efficient `blind' exploration by random sampling.
Our experiments thus suggests that the developers of the tool should consider alternative default choices. 
It also hints that hybridization of the search algorithms with random sampling of a larger stochastic model should be considered in future work.
Given our results it is likely that such a hybrid would make it easier for, for example, the NMCS-based methods to break free from the constraints of their current default model.

Somewhat ironically, but in retrospect naturally, the main conclusion is that \textit{there is not likely to be a single best method or search algorithm to use for different types of test diversity needs}; one needs a toolbox of diverse solutions that needs to be tailored to the diversity goal and situation at hand. This is in line with the argument in~\cite{feldt2015broadening} that researchers in search-based software engineering should not only consider the basic evolutionary algorithms but should open up to consider a richer set of search and optimisation solutions. In particular this will be important in real-world software testing where there is a need to repeatedly explore the same test input feature space, for example in regression testing scenarios. There we argue that if a model of the mapping from the feature space to the parameter space is built up front, for example using Gaussian Processes as proposed in~\cite{feldt2015broadening}, it can be exploited in later sessions to more quickly generate a diverse set of test data. Future work should, of course, also investigate more test data generation scenarios and evaluate how ability to find real and seeded faults is affected by test data diversity and the size of the feature space from which it is sampled. 

\section{Conclusions}
\label{sec:Conclusions}

We have described the feature-specific test diversity problem and investigated how it can be solved with different types of search and sampling approaches. After defining 10 different approaches we evaluated them on a test data generation task for a two-dimensional feature space. Results show that a hill climbing search both gave the best coverage of the target area and was the most efficient (per time step) but that random sampling can be surprisingly effective. The empirical results points to several ways in which the investigated approaches can be improved and, possibly, hybridized to address a diverse set of test data diversity needs. In particular we propose that models that map between feature values and the space being searched can help to ensure test diversity in scenarios of frequent re-testing, such as for regression testing.


Our results also have wider implications for search-based software engineering. Random sampling and, in particular, ways to sample to ensure a better spread over the search space, such as with latin hypercube sampling, can be surprisingly effective in creating diversity. We caution other researchers in search-based software engineering to not blindly reach for a standard search procedure like a genetic algorithm. Depending on the goals for the search and the characteristics of the search and features spaces, non-standard or hybrid methods may be needed and should be considered.




%
%

\bibliographystyle{IEEEtran}
\bibliography{llncs}



\end{document}